\documentclass[useAMS,usenatbib,usegraphicx,epsfig]{mn2e}

\usepackage{amssymb}
\usepackage{graphicx}
\usepackage{amsmath}
\usepackage{multirow}

\title[Inferring the CIB Redshift Distribution]{Inferring the Redshift Distribution of the Cosmic Infrared Background\thanks{Based on observations obtained with Planck (http://www.esa.int/Planck), an ESA science mission with instruments and contributions directly funded by ESA Member States, NASA, and Canada.}}

\author[Schmidt et al.]{
\noindent Samuel J.~Schmidt$^{1,2}$\thanks{email: sschmidt@physics.ucdavis.edu}, 
Brice M\'{e}nard$^{3}$\thanks{Alfred P.~Sloan Fellow}, 
Ryan Scranton$^{1}$,
\newauthor Christopher B.~Morrison$^{1,4}$, 
Mubdi Rahman$^{3}$, Andrew M.~Hopkins$^{5}$\\
\noindent $^{1}$Dept. of Physics, University of California, One Shields Ave., Davis, CA 95616\\
\noindent $^{2}$Dept. of Physics and Astronomy, University of Pittsburgh, 3941 O'Hara Street, Pittsburgh, PA, 15260\\
\noindent $^{3}$Dept. of Physics and Astronomy, Johns Hopkins University, 3400 N.~Charles Street, Baltimore, MD, 21218\\
\noindent $^{4}$Argelander-Institut f\"{u}r Astronomie, Auf dem H\"{u}gel 71, D-53121 Bonn, Germany\\
\noindent $^{5}$Australian Astronomical Observatory, PO Box 915, North Ryde, NSW 1670, Australia
}

\date{Draft, \today}

\begin{document}

\maketitle

\begin{abstract}
Cross-correlating the Planck High Frequency Instrument (HFI) maps against quasars from the Sloan Digital Sky Survey (SDSS) DR7, we estimate the intensity distribution of the Cosmic Infrared Background (CIB) over the redshift range $0 < z < 5$. We detect redshift-dependent spatial cross-correlations between the two datasets using the 857, 545 and 353 GHz channels and we obtain upper limits at 217 GHz consistent with expectations. At all frequencies with detectable signal we infer a redshift distribution peaking around $z \sim 1.2$ and find the recovered spectrum to be consistent with emission arising from star forming galaxies.  By assuming simple modified blackbody and Kennicutt relations, we estimate dust and star formation rate density as a function of redshift, finding results consistent with earlier multiwavelength measurements over a large portion of cosmic history.  However, we note that, lacking mid-infrared coverage, we are not able to make an accurate determination of the mean temperature for the dust responsible for the CIB.  Our results demonstrate that clustering-based redshift inference is a valuable tool for measuring the entire evolution history of the cosmic star formation rate from a single and homogeneous dataset.
\end{abstract}


\section{Introduction}

The cosmic infrared background (CIB), long predicted \citep[e.~g.~][and references therein]{Par:67,Bon:86} and first detected by the Cosmic Background Explorer (COBE) satellite \citep[]{Pug:96}, largely consists of UV/optical photons absorbed by dust and re-emitted in the infrared (see \citet[]{Hau:01} and \citet[]{Kas:05} for comprehensive reviews of early CIB work).  As the bulk of the CIB is not currently resolved into individual sources in the far infrared it has thus far been difficult to constrain the number counts and spectral energy distribution (SED) of the sources, let alone their redshift distribution.  The {\em Spitzer Space Telescope} in the near-infrared \citep[]{Wer:04} and {\em Herschel Space Observatory} in the far-infrared \citep[]{Pil:10} have greatly added to what is known about the CIB, albeit on limited areas of the sky due to their fairly small fields of view.  For example, \citet[]{Dol:06} stack sources detected at 24 microns in Spitzer to determine their contribution to the CIB at longer wavelengths where the populations are confusion limited.  They find that the 24 micron sources brighter than 60 $\mu$Jy contribute 92\% of the CIB at 70 microns and 69\% at 160 microns. \citet[]{Beth:12} similarly stack 24 micron sources in the COSMOS and GOODS-N fields as a function of (photometric) redshift.  They find that the resolved sources account for  55-73\% of the CIB at 250-500 $\mu$m.  Thus, even deep targeted observations with limited sky coverage that rely on stacking of higher frequency detections do not fully resolve the CIB, and are additionally susceptible to sample variance in the limited areal coverage. More recently the release of the Planck High Frequency Instrument (HFI) maps have provided a full sky view of the CIB along the lines of COBE but with much greater resolution, albeit not high enough to resolve individual sources.  To extract the redshift distribution from this
map, we will need an alternative approach.

The use of clustering information to infer redshift distributions is becoming a well established tool in cosmology.  \citet[]{Sch:06} and \citet[]{New:08} described a method using cross-correlations of samples with unknown redshift distributions against photometric and spectroscopic datasets, respectively.  Further work has been proposed for spectroscopic datasets \citep[e.~g.~][]{Matt:10,McQ:13}, though restricted to mock datasets.  Recently, the clustering analysis has been applied using real data \citep[]{Ho:08,Nik:12,Men:13,Rah:14}.  In fact, \citet[]{Mit:12} use a cross-correlation based technique to measure the redshift distribution of resolved far infrared Herschel Multi-tiered Extragalactic Survey (HerMES), albeit only in five broad redshift bins.   In addition, \citet[]{Ser:14} cross-correlated Luminous Red Galaxies (LRGs) from SDSS-III Data Release Eight (DR8) and data from Planck and IRIS to estimate the bias of CIB galaxies with respect to the dark-matter density field and estimate a mean temperature of $T_d=26$ K for the dust responsible for the CIB emission.

In this paper we extend the method further, applying a clustering-based redshift inference technique to a field of unresolved sources. Using data from the Planck satellite and the SDSS, we will benefit from the availability of a large fraction of the sky, therefore reducing sample variance effects. The outline of the paper is as follows: in \S\,\ref{method} we present the method; in \S\,\ref{data} we describe the Planck maps and spectroscopic quasar datasets; in \S\,\ref{results} we present the intensity measurements, as well as estimates of the dust mass density and star formation rate density as a function of redshift; in \S\,\ref{conclusions} we offer conclusions and future directions; and finally, in an Appendix we summarize other findings from Planck maps that may be of interest.  We assume a cosmology with $\Omega_{M}=0.274$, $\Omega_{\Lambda}=0.726$, and $H_{0}=70.5$ km/s/Mpc \citep[]{Kom:09}.  All error bars on the distributions are computed via spatial jackknife.

\section{Method}\label{method}

Our methodology is based on ideas from \citet[]{Matt:10}, \citet[]{Sch:13} and \citet[]{Men:13} (we refer the reader to those papers for full details). Qualitatively, we are using the fact that the sources whose redshift distribution we want to recover are embedded in the same large scale structure as a reference set of objects with known redshifts.  This shared structure leads to a detectable spatial correlation signal when the unknown and reference samples overlap in redshift, but no signal when they do not, enabling us to trace out the redshift distribution by cross-correlating the unknown sample against bins of objects with spectroscopic redshifts.

The two point cross-correlation function, $w(\theta)$, is a measure of the degree to which two datasets are correlated.  We will attempt to recover the redshift intensity distribution of an ``unknown'' sample by cross-correlating with a ``reference'' sample that consists of objects with measured spectroscopic redshifts.  If the unknown and reference samples overlap in redshift, then the associations due to the shared structure will be evident as a positive cross-correlation, whereas if they are physically unassociated then the cross-correlation signal will drop to the Poisson level expected for a random sample.  By dividing the reference spectroscopic sample into redshift bins and cross-correlating each bin with the unknown dataset, we can trace out the redshift distribution: the amount of overlap between the reference spectroscopic and the unknown sample will determine the amplitude of the cross-correlation signal. The measured set of angular cross-correlations provides us with constraints on the product of the redshift distribution of the unknown sample in each redshift bin, ${\rm d N_u/d} z$$(z_{i})$ and the clustering amplitudes of the unknown and reference samples, $b_{u}$ and $b_{r}$, namely:
\begin{equation}
w_{ur}(z_{i})\propto\frac{dN_{u}}{dz}(z_{i})b_{u}(z_{i})b_{r}(z_{i})w_{DM}(z_{i})
\end{equation}\label{dndzeqn}
\noindent where $w_{DM}$ is the expected Dark Matter clustering signal.  We can measure the clustering amplitude of the reference sample in each redshift bin from the reference sample autocorrelation function.  Using a iterative method very similar to that described in \citet[]{Matt:10} we estimate $b_{u}$ for the unknown sample, enabling us to infer ${\rm d N_u/d} z$ from $\bar{w}_{ur}(z_{i})$.

In this paper all pixelization, map manipulation, and cross-correlation measurements are done with the {\rm ASTRO-STOMP} software package\footnote{stomp is available at https://code.google.com/p/astro-stomp/}, hereafter referred to as  ``stomp''.  We measure the set of angular cross-correlation functions between a reference spectroscopic sample
 $r$ and the unknown sample $u$ considering a fixed physical aperture, defined by $\rm{r_{min}}$ and $\rm{r_{max}}$ (in Mpc), which corresponds to angular scale $\rm{\theta_{min}}$ and $\rm {\theta_{max}}$.
We combine the clustering information from all considered scales by integrating the angular correlation according to
\begin{equation}
  \hat{w}_{ur}(z) = \int_{\theta_{min}}^{\theta_{max}}W(\theta')w_{ur}(\theta',z) d\theta'\;.
\end{equation}
As in \citet[]{Sch:13} and \citet[]{Men:13} we use an inverse weight $W(\theta)\propto \theta^{-1}$ normalized over the aperture.   We use an aperture between physical scales of $\rm{r_{min}=0.3\, Mpc}$ and $\rm{r_{max}=3.0 \, Mpc}$, corresponding to $\sim 1-10$ arcminute scales over the redshift interval considered in the paper.
 The intermediate/quasi-linear scales employed in this paper should not be strongly sensitive to the differing bias evolution, as we will show in Section~\ref{results}.
Finally, as done in \citet[]{Sch:13}, we attempt to minimize the effect of the redshift evolution of unknown population bias using the iterative bias correction method similar to that described in \citet[]{Matt:10}.

Unlike previous implementations of cross-correlation methods, in this paper we are no longer measuring the correlation of discrete objects. As the cosmic infrared background is unresolved, our redshift inference technique is not based on source number counts but flux or intensity measurements:
\begin{equation}
\hat{\frac{dI_{\nu,i}}{dz}} = \frac{dI_{\nu,i}}{dz}\bigg/ \int{\frac{dI_{\nu,i}}{dz'}dz' }\;.
\label{inten_eqn}
\end{equation}

In terms of the code machinery used in the calculations, this generalization to continuous distributions is essentially no different than the method used for discrete correlations, now treating weighted pixels rather than individual galaxies.  To test the code, we pixelized the MegaZ-LRG galaxy catalog \citep[]{Col:07} to create a galaxy surface density map on a regular grid of stomp pixels.  The redshift distribution of this sample was previously calculated by cross-correlating with the SDSS quasar catalog in \citet[]{Men:13}, and we successfully recover the expected redshift distribution with the new estimator.

\begin{table*}
\begin{center}
\caption{Total Cosmic Infrared Background Intensity}
\begin{tabular}{|c|c|c|c|c|}
\hline
\multirow{2}{*}{Frequency} & intensity (DR7) within & intensity (NGCAP) within  & Planck VIII monopole & Planck XXX monopole\\
    & 0.3-3 Mpc annuli & 0.3-3 Mpc annuli & & \\
\ [GHz] & [MJy/sr] & [MJy/sr] & [MJy/sr] & [MJy/sr]\\
\hline
217 & 0.0631 $\pm$ 0.0071 & 0.0578$\pm$ 0.0070& $0.033 \pm 0.0066$ & $0.035 \pm 0.0014$\\
353 & 0.177 $\pm$ 0.027 & 0.159 $\pm$ 0.027& $0.13 \pm 0.026$ & $0.150 \pm 0.0057$\\
545 & 0.434 $\pm$ 0.072 & 0.377 $\pm$ 0.072& $0.35 \pm 0.07$ & $0.422 \pm 0.018$\\
857 & 1.09 $\pm$ 0.12 & 0.90 $\pm$ 0.12& $0.64 \pm 0.12$ & $0.898 \pm 0.023$\\
\label{monopole_table}
\end{tabular}
\end{center} 
\end{table*}

\section{Data and Analysis}\label{data}

The spectroscopic reference sample used in the cross-correlation analysis consists of quasars selected from Sloan Digital Sky Survey (SDSS) DR7 \citep[]{Schqso:10}. Areas of bad seeing and high galactic reddening are masked in a manner described by \citet[]{Scr:02}.  Additionally, a 60 arc second radius is excluded around galaxies brighter than $r<16$ and stars with saturated pixels, as is discussed in \citet[]{Scr:05}.  \citet[]{Men:10} used these same seeing, reddening, and bright star masks for measuring the cross-correlation of SDSS photometric quasars and the density of foreground galaxies, and found no significant systematics in the quasar cross-correlation functions.  The spectroscopic quasars used in our analysis are brighter, and thus should be subject to fewer systematic biases than those used in \citet[]{Men:10}.  The DR7 footprint, once these masks have been applied, covers 5364.5 square degrees.  After masking, our sample contains 63995 quasars.  The broad redshift coverage of this reference sample enables us to probe the CIB intensity over a large fraction of cosmic time.

We use the Planck High Frequency Instrument (HFI) maps with Zodiacal light subtracted\footnote{Planck maps are available at: http://irsa.ipac.caltech.edu/data/Planck/release\_1/all-sky-maps/}.   The Planck pixels are $\sim\,1.7\arcmin\,\times\,1.7\arcmin$.  To perform the measurements, we resample the maps from native HEALPix \citep[]{Gor:05} format to create maps in the stomp format, which uses the $\lambda - \eta$ coordinates of the SDSS.  The HFI maps are supplied in units of Kelvin (217 GHz and 353 GHz), or MegaJansky per steradian (MJy/sr, 545 GHz and 857 GHz).  The Planck documentation states that converting MJy/sr to Kelvin can be problematic for the highest frequencies, so we work in MJy/sr and use the conversions listed in Table 6 of Planck 2013 IX \citep[]{planckix:13} to convert the lower frequency maps, namely factors of 483.69 and 287.45 $\rm{MJy\, sr^{-1}\, K^{-1}}$ for the 217 GHz and 353 GHz bands respectively.  

A ``colour correction'' term is also necessary in order to put the intensities in ${\rm \nu I_{\nu}=cst}$ units (similar to AB magnitudes, i.~e.~constant intensity per logarithmic unit in frequency, so a spectrum with ${\rm I_{\nu} \propto \nu^{-1}}$ would have zero colour correction), which depends on the shape assumed for the spectrum.  The actual intensity is related to the ${\rm \nu I\nu=cst}$ intensity by:
\begin{equation}
I_{\nu_{0}}(actual) = \frac{I_{\nu_{0}}(\nu I_{\nu}=cst)}{cc}\;,
\label{cc_eqn1}
\end{equation}
\noindent where the colour correction is given by:
\begin{equation}
cc = \int{\frac{I_{\nu}}{I_{\nu_{0}}}\tau(\nu)d\nu} \bigg/ \int{\frac{\nu_{0}}{\nu}\tau(\nu)d\nu}
\label{cc_eqn2}
\end{equation}
\noindent and $\nu_{0}$ is the effective central frequency of the passband and $\tau(\nu)$ is the filter transmission as a function of frequency.  As we show in \S\,\ref{results} the CIB intensity distribution is similar to that of a typical starburst galaxy with an SED similar to a modified blackbody in the far infrared.  We use the $10^{11}\,L_{\sun}$ ``starburst'' spectrum of \citet[]{Lag:03} to compute the corrections.  The colour correction will change as a function of redshift for such a source, as the slope of the SED changes as it redshifts through the passbands.  For the four Planck bands considered in this paper the colour correction ranges from $\sim$1.0 to 1.1 in the range ${\rm 0<z<5}$.

\subsection{Galactic Foreground Removal}\label{foreground}

A complication of the CIB measurement is that several strong foreground components also contribute to the intensity at far-infrared wavelengths: Zodiacal light, Galactic dust, CO emission, and at lower frequencies, the cosmic microwave background (CMB).  We use Planck maps where Zodiacal light has been removed \citep[see Planck XIV,][for details ]{planckxiv:13}.  Galactic dust emission dominates the signal of the CIB at the high frequency channels in which we expect to detect the background radiation. Large scale gradients in the foregrounds that vary relatively smoothly over the SDSS footprint add an overall  offset in the mean intensity as a function of position.  As they are physically unassociated, these foreground fluctuations should not correlate with our quasar sample, and thus should only affect the amplitude normalization of the intensity distribution, not the shape of the intensity distribution itself.  However, since we estimate errors with a spatial jackknife technique, the slowly varying foregrounds will bias the overall density on larger scale jackknife regions, increasing the estimated uncertainties on the density estimator.

In order to mitigate the large scale emission due to Galactic foregrounds, we divide the DR7 quasar footprint into 8192 spatial regions ($\sim$0.65 deg$^2$ each) and compute the mean intensity at each frequency in each region.  We then subtract this mean in each region from the original map.  This procedure removes the large scale foreground signals to zeroth order, but should not affect the 0.3-3.0 Mpc scales used in the correlation calculations.  While this procedure removes the large scale gradients, it also removes intensity due to the CIB at larger scales.  The clustering analysis still recovers the (normalized) intensity distribution as a function of redshift; however, we must now determine our intensity normalization from an independent method.

\begin{figure*}
    \centering 
    \includegraphics[width=0.7\hsize]{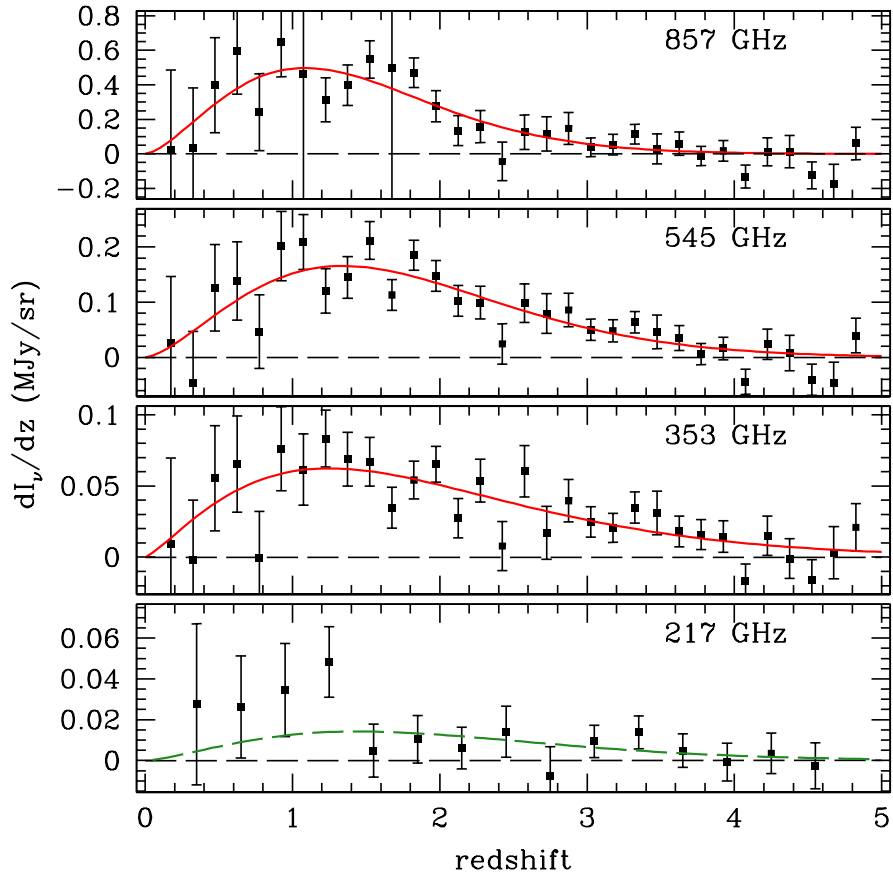}
    \caption{Recovered intensity (in MJy/sr) as a function of redshift at 857, 545, 353, and 217 GHz bands.  The red lines show a parametric fit to the intensity for all but the 217 GHz data; the parameters for the fits are listed in Table~\ref{inten_table}.  For the 217 GHz channel, the predicted strength of the CIB monopole is comparable to the noise level in the Planck map.  The green dashed line indicates the expected signal at that frequency assuming the 545 GHz fit and assuming the $10^{11}L_{\odot}$ Lagache et al spectrum.
    \label{intenfits4}
    }
\end{figure*}

\subsection{Intensity Normalization}\label{normalization}

To determine the overall amplitude of the CIB signal we define a ``Northern Galactic cap'' (NGcap) map, consisting of the portions of each map within the area $\rm{130^{o}<RA<240^{o}}$ and $\rm{5^{o}<DEC<60^{o}}$.  This region, all at or above Galactic latitude $\rm{b\gtrsim25}$, will be less contaminated by Galactic dust emission and other foreground contamination.  To determine the mean CIB intensity, which will be used as the amplitude for our normalized intensity distribution, we sum the intensity in all pixels contained in the 0.3-3.0 Mpc apertures for quasars in the NGcap footprint and divide by the total area of the summed apertures to obtain a mean intensity value (in MJy/steradian).  Table ~\ref{monopole_table} shows our estimates of the CIB monopole both in the entire masked DR7 footprint and in the NGcap section of the map.  Uncertainties are estimated via a spatial jackknife, and the uncertainties due to the zero level and CIB monopole are added in quadrature.  Shown for comparison are the monopole estimates from Planck VIII \citep[]{planckviii:13} and Planck XXX \citep[]{planckxxx:13}, where the latter values are estimated from the model of \citet[]{Beth:12b}.  When computing the CIB monopole, we are systematically high in the DR7 footprint, which is almost certainly due to foreground contamination.  Restricting to the NGcap region shows better agreement with current estimates from Planck VIII \citep[]{planckviii:13}, though the 217 GHz intensity is greater by 65\%.  The 217 GHz band contains a prominent CO line, and thus may be particularly affected by Galactic CO emission.  The normalization in each band will directly affect the flux ratios discussed in the following section, and any biases present would affect our conclusions.  As measurements of the CIB total intensity at these frequencies are still uncertain in the literature (e.~g.~the 857 GHz values in Table~\ref{monopole_table}), we will use our NGcap normalization for the remainder of this paper.

\subsection{Clustering Bias Corrections}\label{biascorr}

The cross-correlation technique actually measures a product of the redshift distribution of the unknown population and its clustering bias. \citet[]{Matt:10} implemented an iterative correction using the autocorrelation functions of both the spectroscopic and photometric datasets to remove the effects of bias evolution in the linear regime.  As shown in \citet[]{Sch:13} using numerical simulations, this technique can be extended into the quasi-linear regime with predictable effects on the accuracy of the recovery.  We will apply this technique to the current analysis.

To estimate the clustering amplitude redshift dependence for the reference sample, we use a fit to the values of the quasar clustering length from \citet[]{Shen:07} and \citet[]{Por:06}.  For the Planck band autocorrelations, we fit power laws to the $C_{\ell}$ values in Table D1 of Planck XXX \citep[]{planckxxx:13} and transform from  ${\ell}$ space to real space using:
\begin{equation}
w(\theta) = \int \frac{k dk}{2\pi}C_{\ell}(k)J_{0}(k \theta)
\label{cltransform}
\end{equation}
where $J_{0}$ is the zeroth order Bessel function and ${\rm k = \ell + 1/2}$, to obtain power law fits for the frequency band autocorrelation functions.  Assuming a form $w(\theta) \propto \theta^{1-\gamma}$, we find values of $\gamma=$ 2.01, 1.82, 1.71, and 1.54 for the 217 GHz, 353 GHz, 545 GHz, and 857 GHz bands respectively.  The values in Table D1 are not corrected for contributions from the thermal Sunyaev-Zeldovich effect and the contamination of the CIB by the CMB, which may be a concern.  However, fits to the clustering length values using Table D2 of \citet[]{planckxxx:13}, which does correct for these effects, produce values that are only 2-3 per cent lower than the $\gamma$ values listed above.  This small difference produces a negligible effect on the inferred redshift distribution, showing that contributions at these frequencies should not affect our clustering measurements.

\section{Results}\label{results}

With our reference and CIB maps properly characterized, we can begin the process of extracting the intensity redshift distribution of the CIB from the Planck maps and using it to infer both the dust mass and star formation rate density histories.

\subsection{An Estimate of the CIB Redshift Distribution}\label{distribution}
Figure~\ref{intenfits4} shows the resulting estimate of the CIB intensity (in MJy/steradian) as a function of redshift for the 217 GHz, 353 GHz, 545 GHz, and 857 GHz bands in bins of width {\rm $\Delta\,z\,=\,0.15$}.  We see a significant coherent signal in three of these bands, recovering broad redshift distributions peaking at $z\sim1.2$.   The expected CIB signal at 217 GHz is near the noise present in the Planck map: the square root of the mean covariance of the 217 GHz map within the DR7 quasar footprint is 1.199$\times10^{-4}\,$K, which translates to a flux intensity uncertainty of 0.058 MJy/sr.  The expected CIB monopole at 217 GHz is 0.033 MJy/sr (Planck VIII 2013, Table 4), lower than the expected noise in the map.  The non-detection in this channel is therefore consistent with expectations.

To characterize the overall redshift distribution we fit a functional form to the data at each frequency:
\begin{equation}
  I_{\nu}(z) = A\, z^{\alpha}\,\rm{exp}\bigg({-\left(\frac{z}{z_{0}}\right)}^{\alpha}\bigg)\,. 
  \label{fiteqn}
\end{equation}
This functional form is somewhat arbitrary. However, it provides us with adequate fits to the distributions given the size of the error bars.  Table~\ref{inten_table} lists the best fit parameter values, as well as the total detection significance (in terms of $\sigma$) for each distribution.  We do not fit a function to the 217 GHz data. Instead we show the expected intensity of the signal as a green dashed line, assuming the spectrum of a ${\rm 10^{11} L_{\sun}}$ starburst galaxy which we normalize using the 545 GHz measured intensity.

Our cross-correlation method is subject to some uncertainty due to the potential differences in the bias evolution between the unknown and spectroscopic samples.  The method proposed by \citet[]{Matt:10} and used above assumes that the unknown bias is proportional to the bias of the reference sample. An iterative approach then provides us with an estimate of the overall amplitude of the unknown bias. As discussed in \citet[]{Sch:13} this procedure is designed to work on large scales, where the clustering amplitude of a given population is expected to be linearly related to that of the dark matter density field. Its accuracy is higher when the unknown population is located within a narrow redshift range, and the relative bias evolution within the bin is minimal.

\begin{figure}
    \centering 
    \includegraphics[width=0.99\hsize]{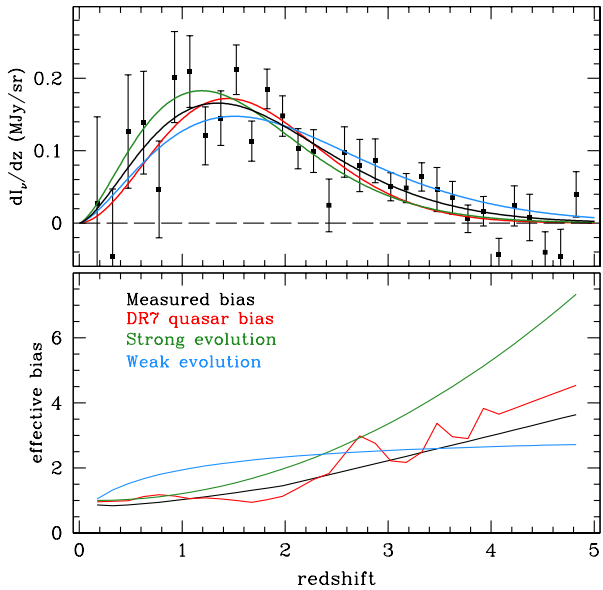}
    \caption{Illustration of the effect of different bias evolution scenarios on the inferred intensity distribution.  The top panel shows the measured intensity distribution in the 545 GHz channel along with different estimates of the recovery for various bias evolution models.   These bias evolution scenarios are shown in the bottom panel, normalized to unity at z=0.15.  While there is some distortion of the recovered shape, uncertainty in the bias evolution is not our dominant source of error.\label{biascompare}
    }
\end{figure}

To estimate the potential uncertainties that could arise due to differing bias evolution, we calculate the distributions given three alternative bias scenarios and compare them to the fiducial result.  The top panel of Figure~\ref{biascompare} shows the parameterized fits for the 545 GHz data from the iterative technique in black, while the bottom shows the evolution of the effective bias (note: the effective bias in all cases has been normalized to unity at z=0.15 for ease of comparison).  The red curve is the result when using a more detailed measurement of the SDSS DR7 quasar bias evolution, which is be presented in \citet[]{Rah:14}.  The green curve is the result for a strongly evolving bias (a quadratic that reaches twice the bias of the nominal case at z=5), and the blue curve for a weakly evolving bias.  We see some distortion in the overall shape and location of the peak intensity of the CIB signal, but the range of values lie mainly within the 1$\sigma$ error bars for the distribution.  Thus, while there is some uncertainty in the overall shape of the distribution, the uncertainty in the bias evolution is not expected to be a strong limitation in inferring the redshift distribution of the CIB.

\begin{figure}
    \centering 
    \includegraphics[width=0.99\hsize]{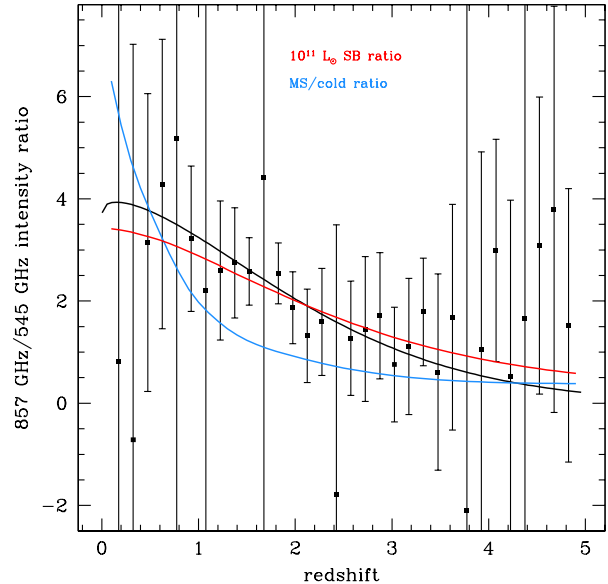}
    \caption{The ratio of 857 GHz to 545 GHz intensities as a function of redshift.  The ratio for the analytic fit is shown in black and this is compared to the ratios calculated from Lagache et al (2003) for Main Sequence/cold (blue) and $10^{11}\,L_{\odot}$ starburst models.  While noisy, the intensity ratios show some preference for the starburst SED.\label{rat3}
    }
\end{figure}

The ratio of the fluxes between the different bands gives us a measure of the mean CIB spectral shape as a function of redshift.  We compute the expected intensity ratios in the Planck bands for the ${\rm 10^{11} L_{\sun}}$ starburst and Main Sequence/cold SEDs from \citet[]{Lag:03}, and compare them to our data.  Figure~\ref{rat3} shows the ratios of 857 GHz to 545 GHz intensities as a function of redshift.  The 545 to 353 GHz and 857 to 353 GHz ratios show very similar behavior. The $10^{11} L_{\sun}$ starburst fits the intensity ratios rather well, which is not unexpected, as the bulk of the CIB at far infrared wavelengths has been thought to arise from just such galaxies \citep[]{Beth:12}.  However, the lack of frequency coverage makes it difficult to constrain the full SED beyond the region of the Rayleigh-Jeans tail of the modified blackbody.  Additionally, without coverage of the modified blackbody peak, it is difficult to put constraints on the mean temperature of the dust that is emitting.  This uncertainty in temperature will propagate to uncertainty in the inferred dust mass density, as we will see in Section~\ref{dustmass}.

Finally, we perform a series of additional measurements to test for potential systematic effects. Similarly to applying our technique to the Planck maps for each channel, we apply it to other types of map: the carbon monoxide (CO) foreground map, the Spectral Matching Independent Component Analysis (SMICA) best extracted CMB map, and the galactic foreground dust opacity map.  For the CO and SMICA maps, we expect no redshift dependent signal, as the signal is expected to originate only from our Galaxy. As shown in the appendix, cross-correlating these maps with SDSS quasars leads to no detectable signal across the entire redshift range available. In contrast, when using the Planck dust map we do detect a redshift-dependent signal. This is expected as it is difficult to disentangle a signal originating from our Galaxy or higher redshifts. Over the 217-857 GHz range, the corresponding spectrum maps onto the featureless power law shape of the Rayleigh-Jeans energy distribution. On performing the measurement, we do see that the intensity as a function of redshift for the dust opacity map is very similar to that which we recover for the CIB signal. As these recovered Planck distributions may be of general interest, we include more detail and show the recovered distributions in the Appendix.\\

\begin{table}
\begin{center}
\caption{Intensity distribution fit parameters}
\begin{tabular}{|c|c|c|c|c|}
\hline
Frequency & A &$z_{0}$ & $\alpha$ & Total S/N\\
\ [GHz] & [MJy/sr] & $-$ & $-$ & [$\sigma$] \\
\hline
217 &   & {\rm noise dominated} &  & 0.91 \\
353 & 0.129  & 1.25 & 1.24 & 4.7 \\
545 & 0.289  & 1.34  & 1.50 & 7.3 \\
857 & 1.19  & 1.08  & 1.52 & 4.2 \\
\label{inten_table}
\end{tabular}
\end{center} 
\end{table}

\begin{figure*}
    \centering 
    \includegraphics[width=0.45\hsize]{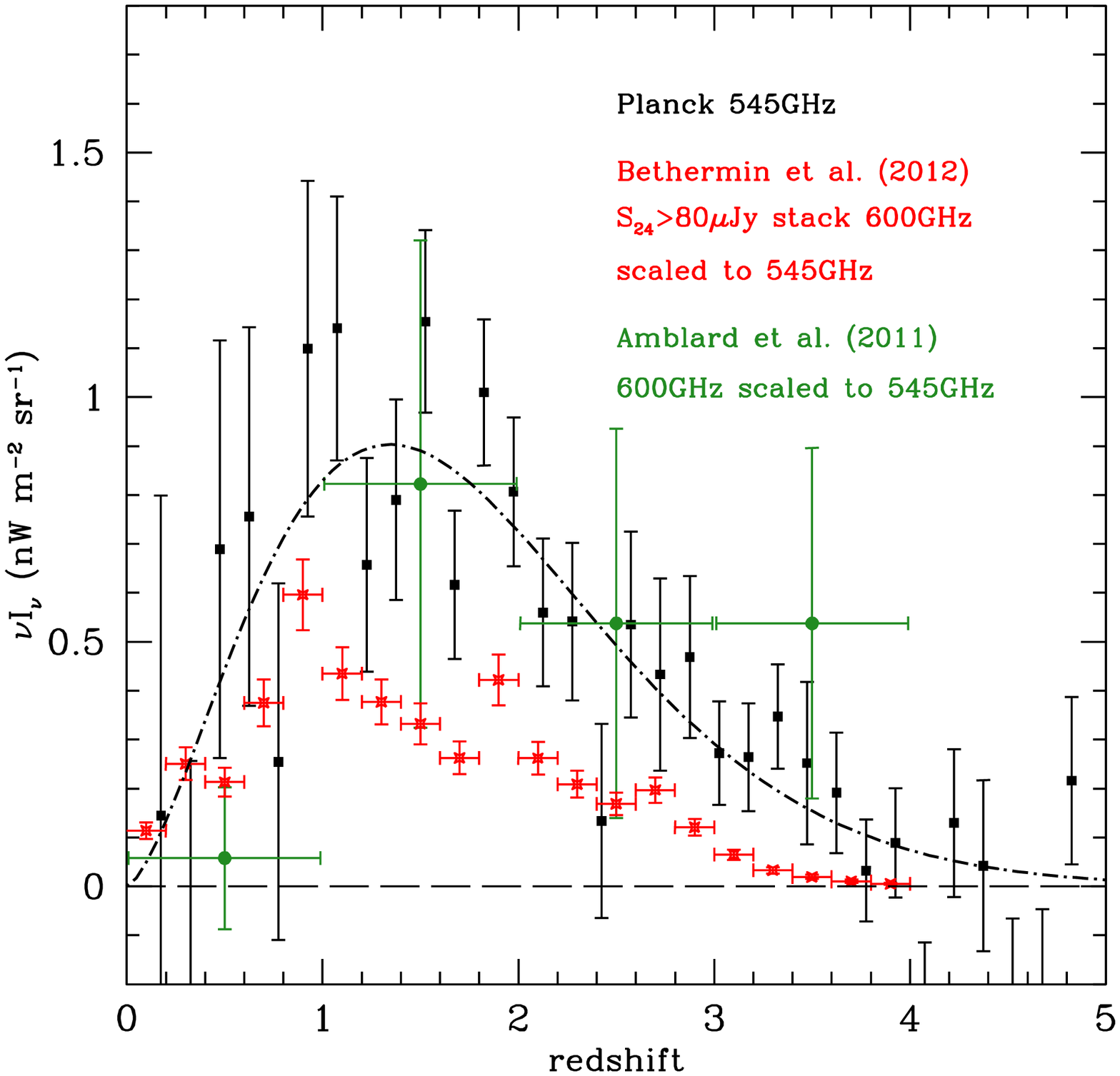}\hspace{0.5cm}\includegraphics[width=0.45\hsize]{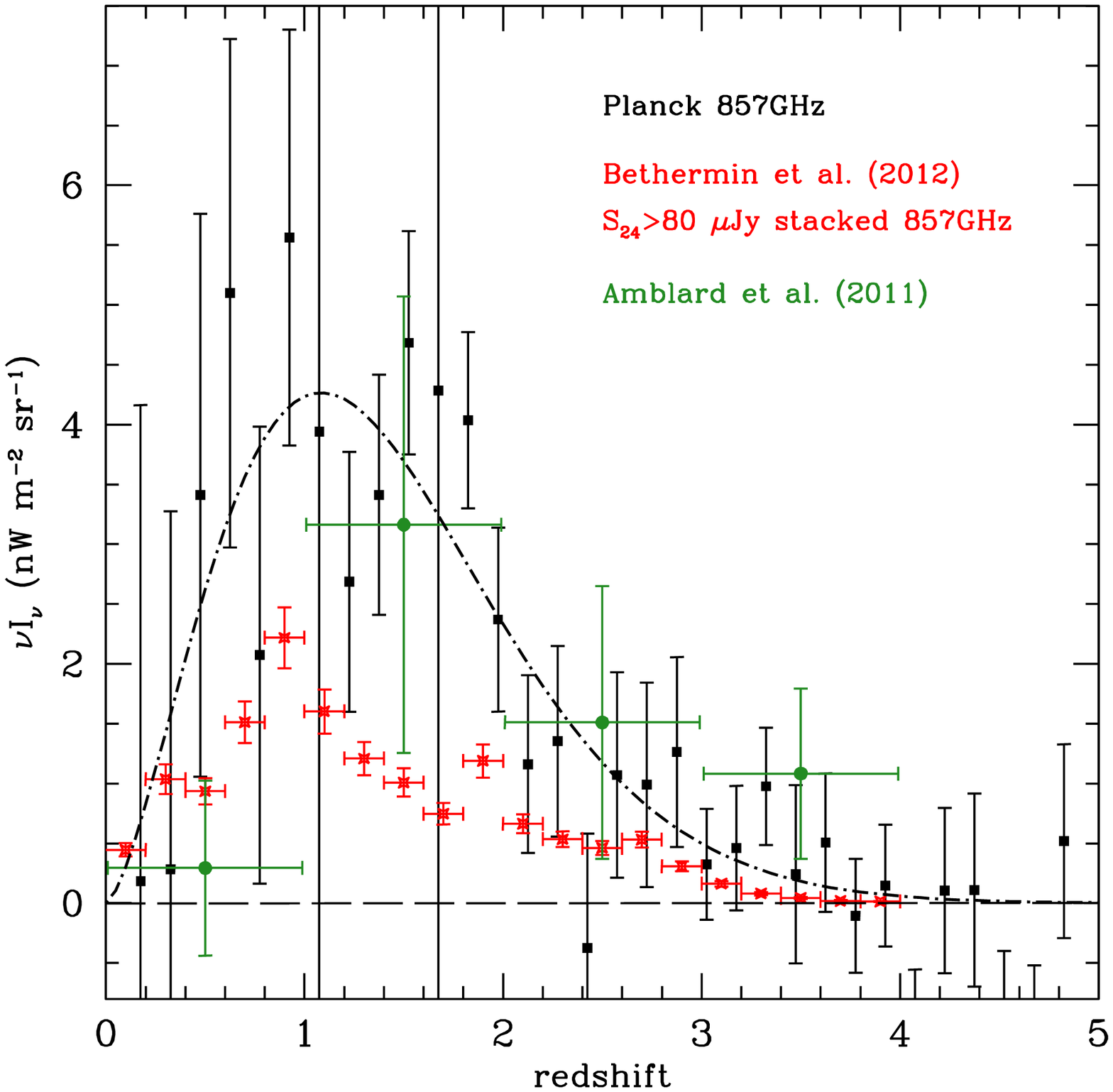}
    \caption{A comparison between the 545 GHz (left) and 857 GHz (right) intensity distribution from Planck (black squares) and the B\'{e}thermin et al.~(2012a) stacked intensity distributions (red squares) and the halo model fit data of \citet[]{Amb:11} (green circles).  In the left panel, data for B\'{e}thermin et al.~and \citet[]{Amb:11} have been multiplied by a factor of 0.70 to scale their 600 GHz observations to match our 545 GHz observations.  We are roughly consistent with the values expected from \citet[]{Amb:11}. \label{nuinu857}
    }
\end{figure*}

\subsection{Comparison to Earlier Results}\label{compare}

There have been multiple efforts in recent years to estimate the redshift distribution of the CIB. \citet[]{Vie:09} use the Balloon-borne Large Aperture Sub-millimeter Telescope (BLAST) clustering signal and a halo model to estimate the CIB signal at 250, 350, and 500 $\mu$m, and find that it arises mainly from the redshift ranges $1.3\,\leq\,z\,\leq\,2.2$, $1.5\,\leq\,z\,\leq\,2.7$, and $1.7\,\leq\,z\,\leq\,3.2$ in each band respectively.  Their model also suggests that star-forming galaxies may reside in the outskirts of groups and clusters at higher redshifts.  \citet[]{Pas:09} use a stacking analysis of BLAST sources detected in Spitzer imaging to estimate the CIB and dust mass using photometric redshifts in six bins, finding that the average dust temperature evolves with redshift, which we explore in the following section.
\citet[]{Mit:12} use a similar cross-correlation based technique to the one employed in this paper, recovering the redshift distribution (not the intensity distribution, as measured in this paper) of resolved HerMES sources with flux density $>$20 mJy by correlating against five broad redshift samples consisting of both spectroscopic and photometric redshifts from the NOAO Deep Wide Field Survey (NDWFS), Spitzer Deep Wide Field Survey (SDWFS), and SDSS.  They find redshift distributions broadly consistent with models of \citet[]{Beth:11}; however, their measurements are affected by some uncertainties. In part this is a matter of small sample size, but there is also a dilution of the clustering signal due to the very large redshift bins employed for the correlations.

We compare our results to those of \citet[]{Beth:12}, who stack sources with {\rm $S_{24\mu\!m} > 80\mu\!Jy$} to estimate their contribution to the CIB at 600GHz and 857GHz.  They use photometric redshifts from COSMOS \citep[]{Ilb:09} to examine the flux intensity distribution as a function of redshift.  Figure~\ref{nuinu857} shows the comparison of the two measurements, where we have converted to units of $\rm{nW/(m^{2}\, sr)}$.
We find markedly higher overall flux intensity than the HerMES values.  However, the 24 micron stacks do not fully resolve the CIB, so the \citet[]{Beth:12} data are only sensitive to a fraction of the overall CIB signal.  \citet[]{Beth:13} show that a $S_{24}>80\,\mu Jy$ selection misses roughly half of the intensity expected from all galaxies.  Including this incompleteness in the \citet[]{Beth:12} results eliminates nearly all of the discrepancy.  In Figure~\ref{nuinu857} we also compare to the data computed from \citet[]{Amb:11}, who use a halo model fit to estimate the mean intensity in four broad redshift bins, which are shown as green circles.  We are roughly consistent with the values at both 545 GHz and 857 GHz.  We note that our CIB monopole determination is slightly higher than expected, so our measurements may need to be scaled down by a small factor due to residual foreground contamination.  

Overall, the shape of the redshift distributions is qualitatively similar to \citet[]{Beth:12}.  We do not see the anomalous peaks at $z\,\approx\,0.3$ and $z\,\approx\,1.9$ seen in both HERMES bands that are attributed to large scale structure in the COSMOS field, however our error bars also do not allow us to rule them out.  Recent papers \citep[e.~g.~][]{Vie:13,Beth:13} have developed models that show an evolution of the peak redshift for the intensity distribution, with the peak shifting to higher redshifts for longer wavelength bands.  We note that our results measure the redshift dependence directly, and not through an intermediate halo model or abundance matching procedure.  The discrepancy could indicate that the assumptions of the model(s) need to be examined further, however, we note that the uncertainties in our measurements near the peak of the redshift distributions are large enough to obscure possible evolution as a function of wavelength.  We also note that we are roughly consistent with the \citet[]{Amb:11} data, which is consistent with the evolving models of both \citet[]{Beth:13} and \citet[]{Vie:13}.  An improved foreground subtraction that further reduces our jackknife uncertainties will enable us to examine this possible discrepancy further, and will be the subject of future work.

\subsection{Dust Mass Density}\label{dustmass}

\begin{figure}
 \centering 
    \includegraphics[width=0.99\hsize]{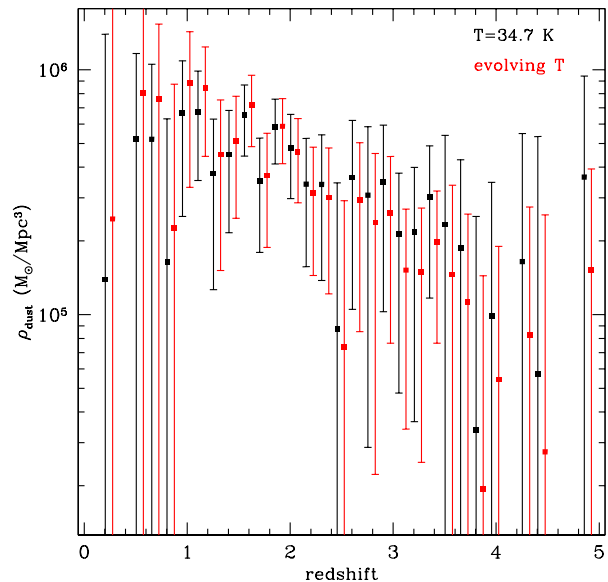}
    \caption{Total dust mass density as a function of redshift using the 545 GHz intensity assuming a constant $T=34.7 K$ (black) and a dust temperature that evolves linearly in ${\rm log(1+z)}$ (red).  While the evolving dust model tilts the shape of the density, values are still within 1$\sigma$ uncertainties, so temperature evolution should not be the dominant factor in the density determination.\label{dustevol}
    }
\end{figure}

Given the unresolved nature of the CIB measurement we cannot estimate the dust mass of individual sources. However, we can estimate the total dust mass density of the aggregate population. To do so, we assume that the dust emission follows a single temperature modified blackbody with emissivity $\epsilon_{\nu}\propto \nu^{\beta}$, which can be described by:
\begin{equation}
I(\nu) \propto \epsilon B(\nu) = A\frac{\nu^{3+ \beta}}{e^\frac{h \nu}{kT_{d}}-1}\;.
\label{mbb}
\end{equation}
\noindent We will fix $\beta = 1.5$ and $T_{d}=34.7$K that best fit the $10^{11}L_{\sun}$ starburst model, though we again caution that this temperature is not well constrained.  Given the intensity, we can estimate the total mass of the dust, 

$M_{d}$, as done in, e.~g.~\citet[]{Mag:12}:
\begin{equation}
M_{d} = \frac{S(\nu_{0}) D_{L}^{2}}{(1+z)\kappa(\nu_{0})B(\nu_{0},T_{d})}
\label{dusteq}
\end{equation}
\noindent where $\rm{B(\nu_{0},T_{d})}$ is the intensity of a blackbody spectrum with temperature $T_{d}$ at $\nu=\nu_{0}$ and $\kappa$ is the dust grain absorption cross section per mass, with the assumed form
\begin{equation}
\kappa = \kappa(\lambda_{0})\left(\frac{\lambda_{0}}{\lambda}\right)^{\beta}\;.
\label{kappaeqn}
\end{equation}
\noindent We use a dust absorption cross section from the revised version of \citet[]{Li:01}\footnote{The revised table with rescaling can be found at http://www.astro.princeton.edu/{\textasciitilde{}}draine/dust/dustmix.html} with ${\rm R_{V}=3.1}$ and $\kappa_{250\mu m} = 5.1 cm^{2}/g$.  We choose the wavelength ${\rm \lambda_{0} = 250 \mu m}$ near the center of our redshifted wavelength coverage to minimize effects of different $\beta$ values for the modified blackbody and cross section dependencies.  These differences can cause systematic bias in dust mass estimates \citep[]{Bia:13} up to factors of several, so some caution must be advised in treating the dust absorption cross section.  To account for this, we have increased our calculated uncertainties by a factor of 2.  Figure~\ref{dustevol} shows the estimated dust mass density, {\rm $\rho_{dust}$}, as a function of redshift for the 545 GHz band (shown in black), found by simply dividing the total dust mass by the volume of the redshift bin.  Results for 353 GHz and 857 GHz are extremely similar.  As our highest frequency band is 857 GHz (350 $\mu$m), we do not have a strong constraint on the dust temperature, as the modified blackbody peaks at $\lambda\,\sim\,75 \mu$m, and all of our data is on the featureless Rayleigh-Jeans tail of the modified blackbody curve at low redshift.  The effective wavelength of the 857 GHz band reaches 75 $\,\mu$m at $z\,\sim\,3.7$; however, larger uncertainties in the measurements once again prevent an accurate estimate of the dust temperature and its possible evolution.  A change in effective temperature can change both the shape and amplitude of the dust mass density as a function of redshift.  

Previous work has found similar difficulty in constraining the dust temperature, particularly with a simple single temperature model.  \citet[]{Dun:11} point out that observed galaxies have multiple phases of dust emission, with warm dust ($\sim$30-40 K) emanating from areas close to energetic heating sources (bright stars), and with colder dust heated by the interstellar radiation field at typically cooler temperatures of 15-20K.  Using Spitzer data spanning $\sim$3-160$\mu$m, \citet[]{Wil:09} fit two temperature dust models and find significant cold ($T\sim18 K$) dust in all types of local spirals, with cold dust comprising a major portion of the IR luminosity. \citet[]{Pas:09} estimate the mean intensity and comoving dust mass density from a stacking analysis of BLAST sources detected in Spitzer using a variety of photometric redshifts in the 0.9 deg$^2$ BLAST GOODS-S Deep field.  They find a strong evolution in the dust temperature with redshift, from T=23.2 K at $0.016\leq\,z\,\leq\,.098$ to T=36.4 K at $1.062\leq\,z\,\leq\,3.5$.  However, they find little evolution in the dust mass density, though they state that dust mass estimates are uncertain at the order of magnitude level for similar reasons to those stated above.  A more sophisticated temperature model would likely revise and improve our dust mass measurements.  

To estimate the effect of evolving dust temperature, we use a simple but reasonable model: a linear (in ${\rm log(1+z)}$) temperature evolution using the values of \citet[]{Pas:09}.  The red points in Figure~\ref{dustevol} show the resulting assumed linear temperature evolution compared to the constant T=34.7 K model.  The dust mass density is tilted relative to the constant temperature model, higher at low redshift and lower at high redshift. However, the values for the evolving model are all within the ample one sigma uncertainties of the constant temperature model, so temperature evolution should not be the dominant uncertainty, if the estimates of \citet[]{Pas:09} are reasonable.  Thus, we conclude that the dust mass density in the Universe has smoothly increased as a function of time, rising roughly an order of magnitude from $z=5$ to the present.  With large uncertainties, the total dust mass density inferred in the local Universe corresponds to $\sim$0.2\% of the total stellar mass density \citep[]{Wilk:08}.  
Intriguingly, this dust mass fraction appears to have declined from a significantly higher value of $\sim$1\% of the total stellar mass density at z$\,\gtrsim\,$2.

\begin{figure}
 \centering 
    \includegraphics[width=0.99\hsize]{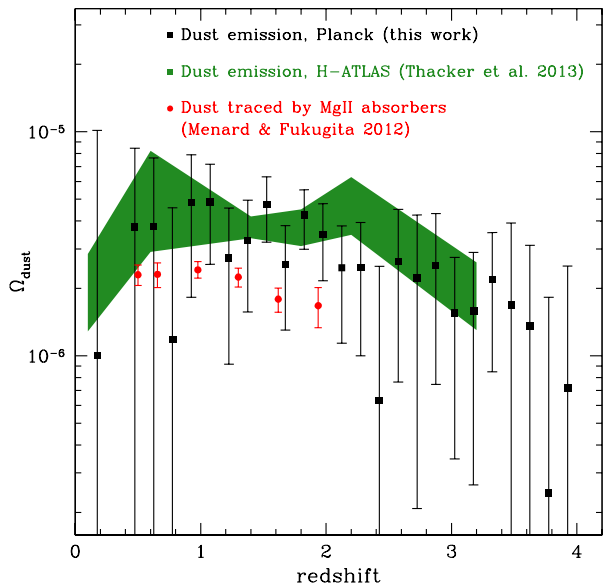}
    \caption{Estimates of dust mass density, $\Omega_{dust}$, as a function of redshift. 
The black data points show our emission-based estimate from spatial cross-correlations between Plank maps and SDSS quasars. 
For comparison we show an estimate of $\Omega_{dust}$ derived from a power spectrum analysis of Herschel observations (Thacker et al.~2013) with the green shaded area. In addition, we show {\rm $\Omega_{\rm dust}$} associated with the circumgalactic medium of galaxies, as probed by MgII absorbers (M\'enard \& Fukugita 2012). This estimate represents a lower limit on the total {\rm $\Omega_{\rm dust}$} value.  Our results are generally consistent with those of the Thacker et al.\label{dustden}
}
\end{figure}

For Figure~\ref{dustden} we have converted our measurement to {\rm $\Omega_{\rm dust}$}, the cosmic dust density in terms of the critical density.  Shown for comparison are two sets of observational results: (i) {\rm $\Omega_{\rm dust}$} estimated from a power spectrum analysis of dust emission mapped out by the Herschel Astrophysical Terahertz Large Area Survey (H-ATLAS) \citep[]{Thack:13}; (ii) {\rm $\Omega_{\rm dust}$} associated with the circumgalactic medium of galaxies, as probed by MgII absorbers \citep[]{Menfu:12}. This estimate is based on dust reddening measurements in the optical. Since the corresponding lines of sight tend to avoid galactic disks, this represents a lower limit on the total {\rm $\Omega_{\rm dust}$} value.  Our measurement is most similar to \citet[]{Thack:13}, and we see general agreement with their results.  This is encouraging, though further investigation is warranted, given our fairly basic assumptions of constant dust temperature and a modified blackbody model.

Our tentative result raises the question of the
balance between dust production, destruction, and dispersion
over the lifetime of a galaxy, and across the breadth of galaxy
populations. There are known trends between star formation
rate and obscuration in galaxies in the local universe \citep[e.~g.~][]{Hop:01,Afon:03,Hop:03,Per:03}. 
The relationship has also been cast as one of increasing obscuration with increasing
stellar mass \citep[e.~g.~][]{Garn:10}. The
SFR-obscuration trend evolves measurably with redshift, but
the trend with stellar mass does not seem to evolve at all at least up
to $z\approx 1.5$ \citep[]{Ly:12,Sob:12,Dom:13}.
The implication seems to be that, while a galaxy is actively
forming stars, dust production is maintained, and at a rate
related to the galaxy's mass.

\begin{figure*}
 \centering 
    \includegraphics[width=0.6\hsize]{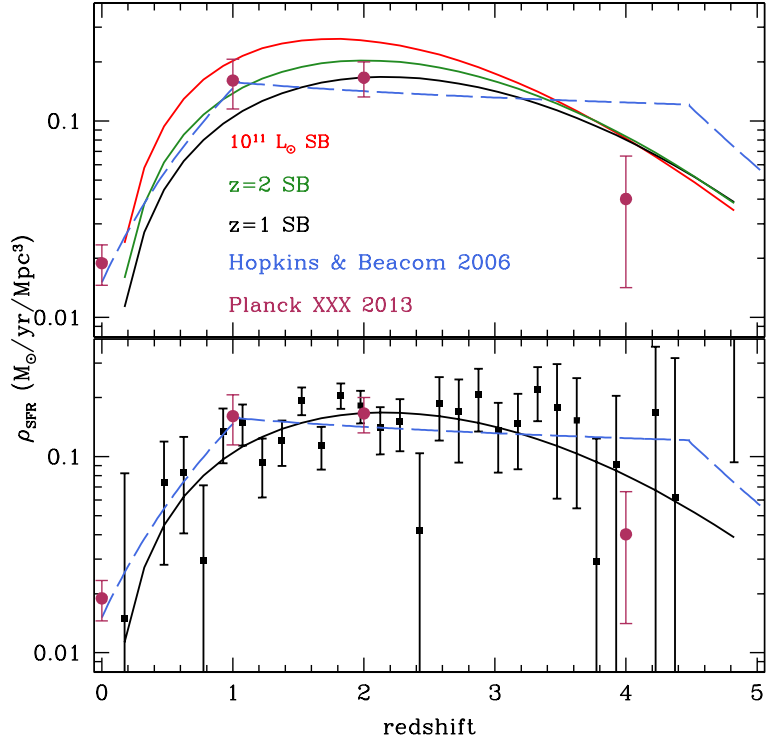}
    \caption{Top: The star formation rate density as a function of redshift for the 545 GHz data, assuming three different SEDs to compute the Bolometric correction: $10^{11} L_{\sun}$ SED from Lagache et al.~(2003) (red), and the empirical SEDs at $z=1$ (black) and $z=2$ (green) of Kirkpatrick et al.~(2012).  The solid lines show the SFR density assuming the analytical fits given in Table~\ref{inten_table}.  The dashed blue line shows the piecewise linear fit to the SFRD from Hopkins \& Beacom (2006), assuming a Salpeter A IMF, and the maroon points show the star formation rate estimate from Planck XXX (2013). Bottom: Same as the top panel, but now showing the computed data points only for the $z=1$ starburst SED.  Our estimate of the SFRD agrees well with the fit of Hopkins \& Beacom.\label{rhosfr}
    }
\end{figure*}

The ``downsizing'' of star formation in galaxies \citep[]{Cow:96} implies that star formation
happens preferentially in lower-mass galaxies at lower redshifts \citep[e.~g.~][]{Mob:09,Jun:05}.
Our observations appear to support this scenario, where
a given star formation rate produces a lower obscuration at lower
redshift, from (on average) a lower mass star forming population.
The evolving fraction of dust mass compared to stellar mass might
be explained through two scenarios: (1) The dust produced is
either destroyed, dispersed or otherwise removed from the galaxy,
at a rate that increases toward lower redshift, such that the
fractional dust content declines; or (2) Stars at lower redshifts
are formed differently, producing less dust per unit stellar mass.
Either or both scenarios (or others) may be operating.
 The downsizing model, combined with recent evidence tentatively indicating a varying
stellar initial mass function \citep[e.~g.~][]{Wilk:08,Gun:11}
 may suggest that a declining level of dust production with decreasing redshift
is a good model to explain the result implied by Figure~\ref{dustevol}.  Once again, decreasing 
the uncertainties in the our $\rho_{dust}$ measurements through improved foreground subtraction or addition of higher frequency bands would enable
better constraints on the evolution of the dust fraction, and will be revisited in future work.


\subsection{The Cosmic Star Formation Rate Density}\label{sfr}

Over the past two decades there have been numerous measurements of the cosmic star formation rate density (SFRD) \citep[e.~g.~][]{Gal:95,Mad:96,Lil:96,Hop:04,Hop:06,Wilk:08,Kis:09,Mcl:13,Kis:13}. As shown below, our results can be used to provide an independent estimate of this quantity.

We can compute the cosmic star formation rate density (SFRD) with a few 
simple assumptions.
First, we convert from intensity $I_{\nu}$(MJy/sr) to spectral flux density $S_{\nu}$(MJy); given our large area ($\sim 5400 {\rm deg}^2$), sample variance should be minimal.  From here, we convert to luminosity $L_{\nu}$ with:
\begin{equation}
L_{\nu(1+z)}=\frac{S_{\nu}4 \pi D_{L}^2}{1+z}
\end{equation}
where $D_{L}$ is the luminosity distance. In order to apply Kennicutt's relation we need 
a bolometric luminosity $L_{Bol}$ estimate obtained by integrating an SED over the range 8-1000 microns (300-37500 GHz). For simplicity, we will assume that the CIB arises from a single, non-evolving, SED template. We compute the $L_{\nu(1+z)}$ to $L_{Bol}$ conversion by fitting our three FIR data points to the template in order to fix the normalization, and simply integrating:
\begin{equation}
  L_{IR}= \int_{8\mu m}^{1000 \mu m}L_{\nu}d\nu\;.
\end{equation}\label{lbol_conv}
The bolometric conversion is very sensitive to the assumed SED over the broad wavelength interval considered. 
Having only access to frequencies lower than 857 GHz limits our ability to constrain the overall SED. The literature presents a range of potential SEDs that give varying bolometric corrections depending on the amount of relative near and mid infrared flux.  Without mid-infrared data to constrain the SED shape it is very difficult to properly estimate the bolometric IR luminosity.  \citet[]{Kirk:12} present empirically derived infrared galaxy SEDs at redshifts $z=1$ and $z=2$ that may be more appropriate than the previously used \citet[]{Lag:03} model.  
These empirical SEDs have a much broader FIR peak, indicating a range of dust temperatures, consistent with references in Section~\ref{dustmass}. There is significantly more flux at mid-infrared wavelengths in these models, resulting in a smaller bolometric correction. 

Once a bolometric luminosity has been computed, we use the relation between SFR and $L_{IR}$ from \citet[]{Kenn:98}:
\begin{equation}
  SFR (M_{\sun}/yr)= 4.5\times\,10^{-44} L_{Bol}(erg/s)\,.
\end{equation}

We compute $\rho_{SFR}$ ($M_{\sun} yr^{-1}Mpc^{-3}$) by dividing the SFR in each slice by the total volume in each redshift bin.  Figure~\ref{rhosfr} shows the star formation rate density computed using the 353, 545, and 857 GHz bands.  The top panel shows the SFRD computed using the analytic fits of Table~\ref{inten_table} using three different SEDs to compute bolometric corrections: the previously used $10^{11} L_{\sun}$ starburst, and the $z=1$ and $z=2$ empirical starbursts of \citet[]{Kirk:12}.  The bottom panel shows the individual bin measurements only for the $z=1$ starburst SED for clarity.  The piecewise-linear fit to the measured total (UV and IR) star formation rate density, assuming a modified Salpeter A initial mass function (IMF) \citep[][Table 2]{Hop:06}, is shown as a dashed blue line for comparison. Uncertainty in the assumed IMF propagates to an uncertainty in the SFRD normalization \citep[]{Hop:06,Wilk:08,Gun:11}. Thus, the true SFRD may be lower by up to a factor of $\sim 2$ compared to what is shown in Figure~\ref{rhosfr}.  Also shown as maroon circles are the SFRD estimates using a linear bias model from \citet[]{planckxxx:13}.  We broadly reproduce SFRD calculated with other methods, with a rise between $0<z<1$, with a plateau and possible decrease at higher redshifts.  The inferred star formation rate at $z>3$ is strongly dependent on the assumed mean SED of the CIB emitting galaxies and their luminosity distribution, which has led to some disagreement in the high redshift SFRD.  We have made no attempt to account for evolution of the mean SED with redshift.  \citet[]{planckxxx:13} compute both a linear bias model and a halo model based estimate of the SFRD and see discrepant values at $z>3$, which can be attributed to the different effective temperatures of the SEDs assumed in the two models.  If the mean temperature of the dust does increase with redshift \citep[e.~g.~][]{Pas:09}, then the $L_{Bol}$ conversion factor of Equation~\ref{lbol_conv} will also evolve and modify the inferred star formation rate density.  A more complex modeling including such effects is left to future work.  It is worth noting that our estimate uses a single data set across the entire redshift range considered.  As such, it is unaffected by cross-calibration issues between estimators based on different tracers and wavelengths.

Given that much of our CIB signal comes from galaxies near $z\sim1$ where the intensity distribution peaks, we note that the \citet[]{Kirk:12}~$z=1$ starburst may be the most appropriate for much of the redshift range.  \citet[]{Chen:13} found that the bolometric correction from 250 $\mu$m Herschel SPIRE observations to  $L_{IR}$ was consistent with the $z=1$ Kirkpatrick starburst SED for a sample of 25 star forming galaxies in the redshift range $0.47\,<\,z\,<\,1.24$, supporting the use of that SED in computing the bolometric correction, at least at $z\,\sim\,1$. No attempt has been made to account for the fraction of UV photons that are not absorbed by dust and escape their host galaxy, which is dependent on at least the mass of the host galaxy.  The total star formation rate density will thus be higher than the FIR star formation rate density shown in Figure~\ref{rhosfr}.  \citet[]{Tak:05} use arguments based on the ultraviolet (UV) and IR galaxy luminosity functions to predict that the fraction of obscured star formation increases with redshift, rising from $\sim$56 per cent at redshift zero to greater than 80 per cent at $z\sim1.2$. This argues that the unobscured star formation not detected by our FIR measurements is a smaller fraction at high redshift.
Once accounting for the additional UV escape photons, our measurement appears to be higher than that observed in the \citet[]{Hop:06} fit.  This is most likely due to the contamination of our CIB signal by foreground dust, though the aforementioned IMF uncertainty may also be a factor.  A more careful treatment of the foregrounds will likely lower the intensity of our CIB amplitude, in turn lowering the SFR measurements.  This is also left to future work.

\section{Conclusions and Future Work}\label{conclusions}

Cross-correlating observations of the microwave sky by the Planck High Frequency Instrument (HFI) against the 3D density distribution of quasars from the Sloan Digital Sky Survey (SDSS) DR7, we characterized the intensity distribution of the Cosmic Infrared Background (CIB) over the range $0 < z < 5$. We detected redshift-dependent spatial cross-correlations between the two datasets using the 857, 545 and 353 GHz channels and obtained upper limits at 217 GHz consistent with expectations. At all frequencies we inferred a redshift distribution peaking around $z \sim 1.2$ and found the recovered spectrum to be consistent with emission arising from star forming galaxies.

Detecting redshift-dependent spatial correlations using Planck observations first requires subtracting the Galactic foreground signal, which is difficult to disentangle from the CIB signal itself in the far-infrared.  We removed the foregrounds at scales larger than those probed in our physical annulus (0.3-3.0 Mpc) by subtracting the mean intensity in spatial regions of $\sim\,0.65\, deg^{2}$ to remove large scale gradients.  As the foreground emission should not cluster with the CIB signal, the residuals should not have affected the recovered redshift distribution, but only increase the uncertainties on the measurement.  Normalizing the overall amplitude of the expected infrared intensity as a function of redshift required an estimate of the CIB monopole.  Our estimate of the total intensity of the CIB signal in the NGCap region tended to agree with monopole measurements in Planck XXX but were systematically higher than those from Planck VIII. This is most likely due to residual signal from the method used for foreground dust subtraction.  A more careful treatment of foregrounds should reduce the uncertainties in the intensity distributions, and is left for future analysis.  

The cross-correlation analysis recovers a product of the intensity distribution and the clustering bias of the underlying datasets.  We assumed a model for the DR7 quasar bias from the clustering lengths given in \citet[]{Shen:07} and \citet[]{Por:06}.  A lack of knowledge of clustering bias evolution for the unknown sample was a potential uncertainty source.  To address this point we considered a range of possible scenarios in Section~\ref{distribution} and showed that this limitation is not expected to be the main source of uncertainty in our estimates.  A more detailed study of the bias properties for several spectroscopic datasets has recently been completed \citep[]{Rah:14}, which may be used in future analysis.

We estimated the total dust mass density as function of redshift in Section~\ref{dustmass}.  With data covering only 217-857 GHz in the observed frame we were unable to place a strong constraint on the dust temperature.  Instead, we fixed the temperature $T_{eff}=34.7 K$ and assumed a modified blackbody with $\beta=1.5$.  We also considered a basic, but empirically motivated, temperature evolution model in Figure~\ref{dustevol}. The resultant change relative to the fixed temperature model was within the measurement uncertainties, so the details in the temperature evolution are unlikely to be the dominant source of error in our dust mass estimates.  Despite these limitations, we observed a trend of gradually increasing dust mass as the Universe ages, with a total increase by almost an order of magnitude since ${\rm z=5}$, which may be consistent with downsizing scenarios.

We used the Kennicutt relation to convert bolometric IR luminosity to the star formation rate density as a function of redshift in Section~\ref{sfr}.  As $L_{IR}$ is dependent on knowledge of the SED at near and mid-infrared wavelengths, our lack of data at frequencies above 857 GHz allowed for some uncertainty in the bolometric correction and, hence, inferred SFR.  As shown in Figure~\ref{rhosfr}, we reproduced the expected qualitative evolution of the SFRD, rising between $0<z<1$, then leveling off or falling at higher redshift. This was true for all three of the different assumed SEDs considered, indicating that this a robust behavior.  These results confirmed our technique as a valuable new tool for measuring the evolution of the cosmic star formation rate from a single dataset.  We are sensitive only to starlight reprocessed into the far infrared, so values shown do not include a correction for UV photons that have not been absorbed and reprocessed into the infrared.  As our estimates of the SFRD are near the mean values measured by \citet[]{Hop:06} before accounting for unobscured star formation in the UV, we believe that our overall intensity for the CIB is overestimated due to the simple method used in the removal of the foreground Galactic dust. 

In this paper we assumed no evolution in the dust temperature and average SED when calculating the dust mass and SFR, which is clearly too simple. Comparison with more complex models (e.~g.~ \citet[]{Lag:04} and \citet[]{Beth:12b}) may differentiate between possible star formation and dust histories.  Adding data at near- and mid-infrared wavelengths spanning the $\sim$70-100 micron peak of the dust emission would provide much better constraints on the SED and effective temperature, enabling better measurements of the dust mass and star formation rate.  As the bulk of the CIB is resolved at shorter wavelengths, surveys such as the Wide-Field Infrared Survey Explorer (WISE) mission \citep[]{WISE:10} may be useful in providing this data, even if such datasets include only the resolved sources. Such an investigation will be the subject of subsequent work.

\section*{Acknowledgements}

SJS was supported by National Science Foundation Grant AST-1009514, Department of Energy Grant DESC0009999, and Department of Energy Early Career program via grant DE-SC0003960. This work is supported by NASA grant 12-ADAP12-0270 and National Science Foundation grant AST-1313302. 
CM is supported by the DFG grant Hi 1495/2-1 and National Science Foundation Grant AST-1009514.  The authors would like to thank the anonymous referee for comments and suggestions that improved the paper.  The authors would also like to thank Russell Ryan and Harry Ferguson for useful discussions and advice early in the course of writing this paper, and Lloyd Knox and Marius Milea for questions on Planck data.  SJS would like to thank Conan O'Brien for twenty plus years of entertainment.

This research has made use of the NASA/IPAC Infrared Science Archive, which is operated by the Jet Propulsion Laboratory, California Institute of Technology, under contract with the National Aeronautics and Space Administration.

The SDSS is managed by the Astrophysical Research Consortium for the Participating Institutions. The Participating Institutions are the American Museum of Natural History, Astrophysical Institute Potsdam, University of Basel, University of Cambridge, Case Western Reserve University, University of Chicago, Drexel University, Fermilab, the Institute for Advanced Study, the Japan Participation Group, Johns Hopkins University, the Joint Institute for Nuclear Astrophysics, the Kavli Institute for Particle Astrophysics and Cosmology, the Korean Scientist Group, the Chinese Academy of Sciences (LAMOST), Los Alamos National Laboratory, the Max-Planck-Institute for Astronomy (MPIA), the Max-Planck-Institute for Astrophysics (MPA), New Mexico State University, Ohio State University, University of Pittsburgh, University of Portsmouth, Princeton University, the United States Naval Observatory, and the University of Washington.

\appendix
\section{Correlation with other Planck maps}

In addition to the dust map, we cross-correlated the quasars with several other Planck map products.  These results are not directly applicable to the CIB, but are interesting systematics tests for Planck data in general, and are presented here for those who may be interested.  

\begin{figure}
\centering 
    \includegraphics[width=0.99\hsize]{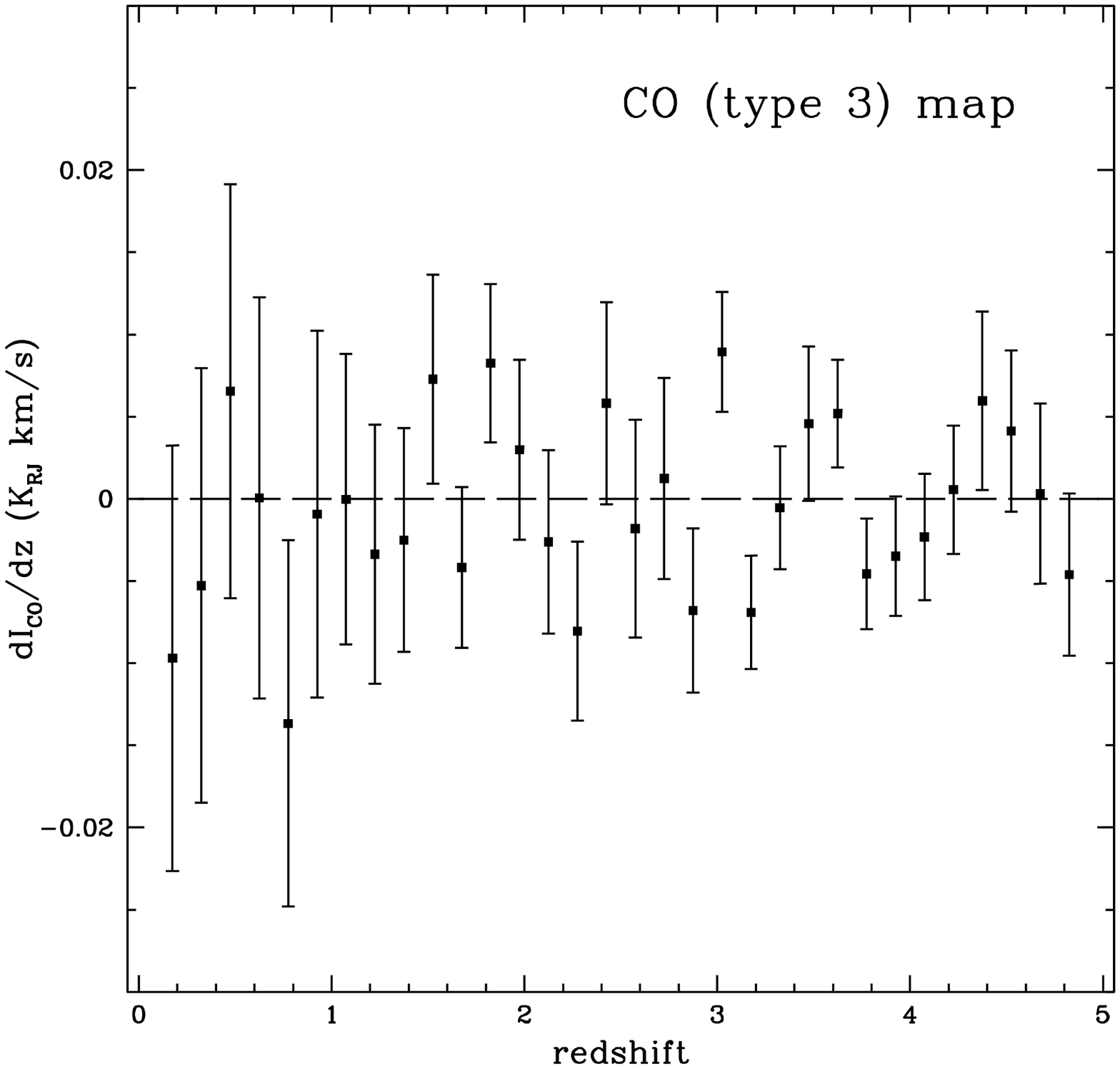}
    \caption{The (type 3) CO intensity distribution as a function of redshift.  No coherent signal is observed, indicating that the CO map is well separated from any extragalactic contamination.
    }
\label{COfig}
\end{figure}

Figure~\ref{COfig} shows the distribution resulting from the cross-correlation of the carbon monoxide (CO) type-3 map and the SDSS quasar sample.  We see no significant correlation with the quasars, indicating a clean separation of the CO signal from any extragalactic contaminants.  As the carbon monoxide transitions are well understood and have distinct spectral characteristics, we expect such a clean separation. 

\begin{figure}
\centering 
    \includegraphics[width=0.99\hsize]{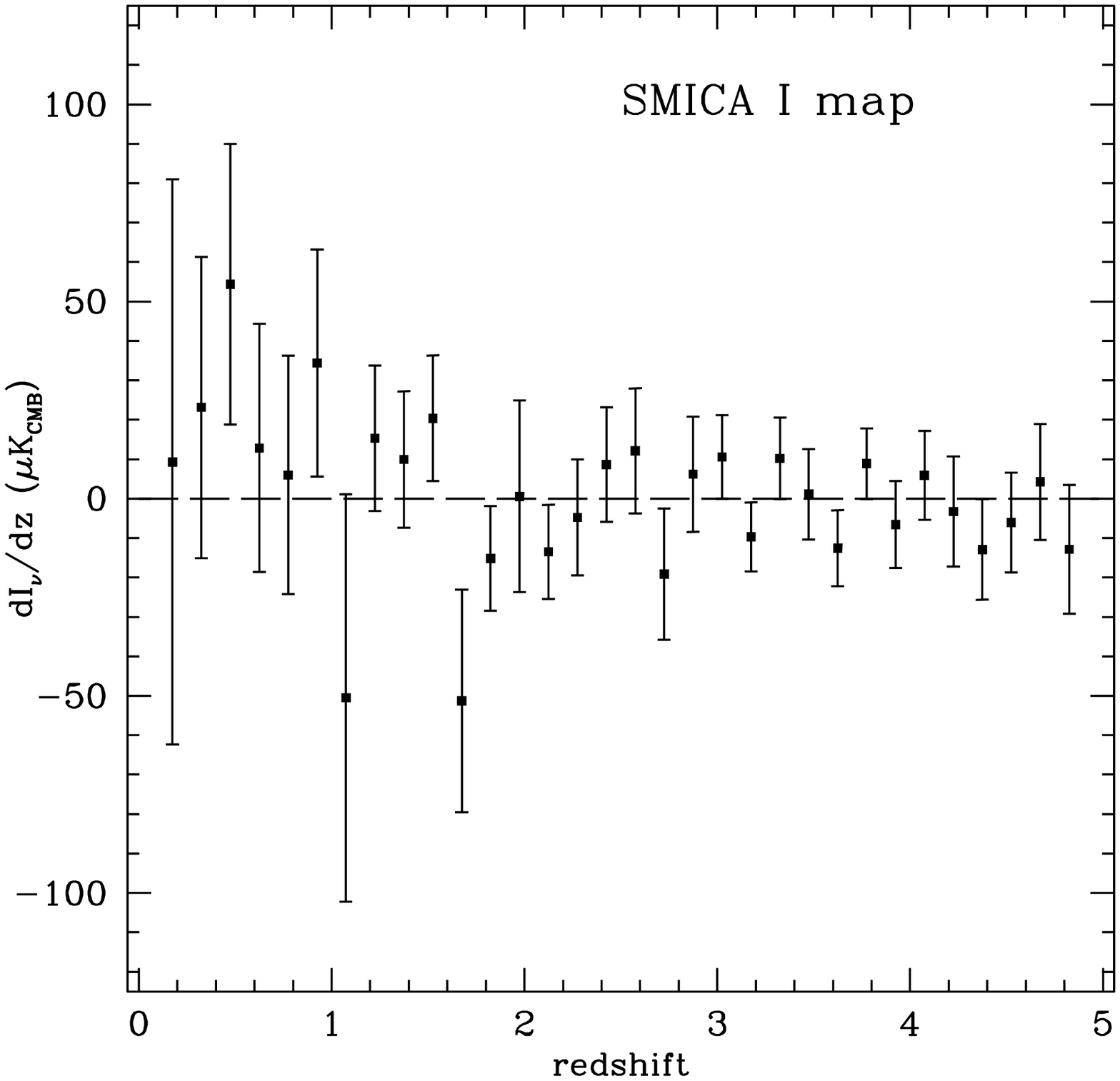}
    \caption{The intensity distribution (in units of ${\rm \mu K_{CMB}}$) for the SMICA map.  No coherent signal is detected at low redshifts and no evidence for low redshift contaminants are seen.
    }
\label{SMICAfig}
\end{figure}

 Figure~\ref{SMICAfig} shows the resulting distribution from cross-correlation of the Planck Spectral Matching Independent Component Analysis (SMICA) map of the CMB emission.  The SMICA algorithm constructs a CMB map from a weighted linear combination of all nine Planck frequency bands as a function of multipole $\ell$ to separate components and isolate the CMB signal (see the Planck XII: Component Separation paper \citet[]{planckxii:13} for details).  While the first six bins all show positive correlation, the level is not significant, and no coherent signal is observed at $z>1$.  Once again, it appears that the CMB component separation has successfully removed extragalactic components, at least to the levels detectable from the cross-correlations presented here.

\begin{figure}
\centering 
    \includegraphics[width=0.99\hsize]{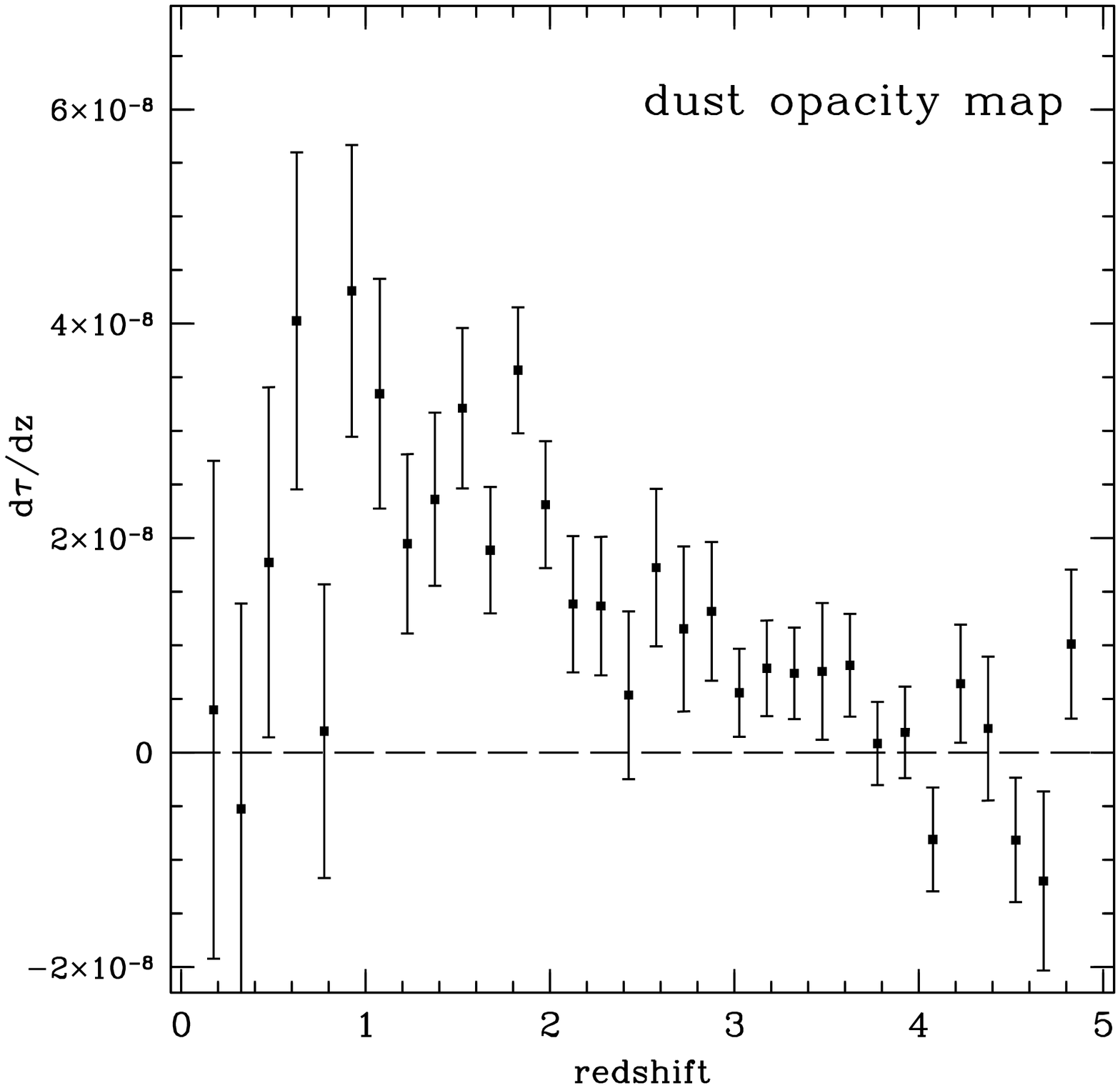}
    \caption{Intensity distribution as a function of redshift for the Planck opacity map cross-correlated with the DR7 quasars, in units of dust opacity, $\tau$.  The significant correlation mimics the redshift and intensity distributions of the CIB, indicating that signal from the CIB galaxies is masquerading as Galactic dust, and that the reddening maps are polluted by this extragalactic signal.
    }
\label{opacityfig}
\end{figure}

As mentioned in \S\,\ref{data}, we found correlations between the foreground dust map and the quasar catalog, strongly suggesting that the foreground dust map was contaminated with the CIB signal that we are measuring.  This is not unexpected: The presence of the CIB is well established, but it is dealt with by subtracting a monopole term, i.~e.~ simply subtracting the expected zero-point amplitude in MJy/sr from the overall map and not taking into account that the CIB is clustered \citep[]{planckviii:13}.  The CIB signal in the far infrared has a very similar spectral shape (similar to a modified blackbody power law in the FIR) to that of Galactic dust at certain temperatures, making it difficult to disentangle the two components.  Thus, some of the small scale power attributed to Galactic dust is actually due to clustered infrared sources.  Figure~\ref{opacityfig} shows the recovered intensity distribution resulting from cross-correlating the SDSS quasar sample with the Planck component map of dust opacity.  We see a signal quite similar in redshift distribution to that of the CIB, reinforcing our assertion that the Planck dust map contains clustered signal from the unresolved galaxies at higher redshifts.  We are working on a method to better disentangle the Galactic and cosmological dust signals.

\section{Tabulated Data}\label{datatables}
This appendix lists the data shown in Figure~\ref{intenfits4} in tabular form.

\begin{table}
\begin{center}
\caption{Cosmic Infrared Background Intensity as presented in Figure~\ref{intenfits4}.}
\begin{tabular}{|c|c|c|c|}
\hline
redshift & 353 GHz Intensity & 545 GHz intensity & 857 GHz intensity\\
 & [MJy/sr] & [MJy/sr] & [MJy/sr]\\
\hline
0.175 & 0.009$\pm$0.060 & 0.027$\pm$0.120 & 0.021$\pm$0.464 \\
0.325 & -0.002$\pm$0.042 & -0.046$\pm$0.093 & 0.033$\pm$0.349 \\
0.475 & 0.055$\pm$0.037 & 0.127$\pm$0.078 & 0.398$\pm$0.275 \\
0.625 & 0.065$\pm$0.034 & 0.139$\pm$0.071 & 0.595$\pm$0.248 \\
0.775 & -0.001$\pm$0.033 & 0.047$\pm$0.067 & 0.242$\pm$0.223 \\
0.925 & 0.076$\pm$0.029 & 0.202$\pm$0.063 & 0.649$\pm$0.203 \\
1.075 & 0.062$\pm$0.025 & 0.209$\pm$0.050 & 0.460$\pm$211.167 \\
1.225 & 0.083$\pm$0.020 & 0.121$\pm$0.040 & 0.313$\pm$0.127 \\
1.375 & 0.069$\pm$0.019 & 0.145$\pm$0.038 & 0.398$\pm$0.117 \\
1.525 & 0.067$\pm$0.017 & 0.212$\pm$0.034 & 0.547$\pm$0.109 \\
1.675 & 0.035$\pm$0.014 & 0.113$\pm$0.028 & 0.500$\pm$76.174 \\
1.825 & 0.054$\pm$0.013 & 0.185$\pm$0.027 & 0.471$\pm$0.086 \\
1.975 & 0.065$\pm$0.013 & 0.148$\pm$0.028 & 0.277$\pm$0.090 \\
2.125 & 0.027$\pm$0.014 & 0.103$\pm$0.028 & 0.136$\pm$0.087 \\
2.275 & 0.054$\pm$0.015 & 0.099$\pm$0.030 & 0.158$\pm$0.093 \\
2.425 & 0.008$\pm$0.017 & 0.025$\pm$0.036 & -0.044$\pm$0.112 \\
2.575 & 0.060$\pm$0.018 & 0.098$\pm$0.035 & 0.125$\pm$0.100 \\
2.725 & 0.017$\pm$0.019 & 0.080$\pm$0.036 & 0.116$\pm$0.100 \\
2.875 & 0.040$\pm$0.015 & 0.086$\pm$0.030 & 0.147$\pm$0.093 \\
3.025 & 0.025$\pm$0.011 & 0.050$\pm$0.019 & 0.038$\pm$0.054 \\
3.175 & 0.021$\pm$0.010 & 0.048$\pm$0.020 & 0.054$\pm$0.061 \\
3.325 & 0.035$\pm$0.011 & 0.064$\pm$0.020 & 0.114$\pm$0.057 \\
3.475 & 0.031$\pm$0.015 & 0.046$\pm$0.030 & 0.028$\pm$0.087 \\
3.625 & 0.018$\pm$0.011 & 0.035$\pm$0.023 & 0.059$\pm$0.067 \\
3.775 & 0.016$\pm$0.011 & 0.006$\pm$0.019 & -0.012$\pm$0.056 \\
3.925 & 0.014$\pm$0.011 & 0.016$\pm$0.020 & 0.017$\pm$0.059 \\
4.075 & -0.017$\pm$0.012 & -0.044$\pm$0.023 & -0.131$\pm$0.067 \\
4.225 & 0.015$\pm$0.014 & 0.024$\pm$0.028 & 0.012$\pm$0.081 \\
4.375 & -0.001$\pm$0.014 & 0.008$\pm$0.032 & 0.013$\pm$0.094 \\
4.525 & -0.016$\pm$0.014 & -0.040$\pm$0.028 & -0.125$\pm$0.078 \\
4.675 & 0.003$\pm$0.018 & -0.046$\pm$0.038 & -0.176$\pm$0.116 \\
4.825 & 0.021$\pm$0.017 & 0.040$\pm$0.031 & 0.060$\pm$0.095 \\
\label{inten_table1}
\end{tabular}
\end{center} 
\end{table}

\begin{table}
\begin{center}
\caption{Cosmic Infrared Background Intensity for 217 GHz as presented in the bottom panel of Figure~\ref{intenfits4}.}
\begin{tabular}{|c|c|}
\hline
redshift & 217 GHz Intensity \\
 & [MJy/sr]\\
\hline
0.35 & 0.476$\pm$0.684 \\
0.65 & 0.454$\pm$0.432 \\
0.95 & 0.598$\pm$0.395 \\
1.25 & 0.835$\pm$0.298 \\
1.55 & 0.084$\pm$0.225 \\
1.85 & 0.179$\pm$0.202 \\
2.15 & 0.106$\pm$0.177 \\
2.45 & 0.244$\pm$0.216 \\
2.75 & -0.131$\pm$0.247 \\
3.05 & 0.163$\pm$0.137 \\
3.35 & 0.239$\pm$0.140 \\
3.65 & 0.083$\pm$0.142 \\
3.95 & -0.014$\pm$0.159 \\
4.25 & 0.060$\pm$0.172 \\
4.55 & -0.045$\pm$0.195 \\
\label{inten_table2}
\end{tabular}
\end{center} 
\end{table}

\bibliographystyle{mn2e}
\bibliography{references}

\end{document}